\documentclass[12pt]{article}
\renewcommand{\theequation}{\arabic{section}.\arabic{equation}}

\begin{document}



\def\a{\alpha}
\def\b{\beta}
\def\d{\delta}
\def\e{\epsilon}
\def\g{\gamma}
\def\h{\mathfrak{h}}
\def\k{\kappa}
\def\l{\lambda}
\def\o{\omega}
\def\p{\wp}
\def\r{\rho}
\def\t{\theta}
\def\s{\sigma}
\def\z{\zeta}
\def\x{\xi}
 \def\A{{\cal{A}}}
 \def\B{{\cal{B}}}
 \def\C{{\cal{C}}}
 \def\D{{\cal{D}}}
\def\G{\Gamma}
\def\K{{\cal{K}}}
\def\O{\Omega}
\def\L{\Lambda}
\def\f{E_{\tau,\eta}(sl_2)}
\def\E{E_{\tau,\eta}(sl_n)}
\def\Zb{\mathbf{Z}}
\def\Cb{\mathcal{C}}

\def\R{\overline{R}}

\def\beq{\begin{equation}}
\def\eeq{\end{equation}}
\def\bea{\begin{eqnarray}}
\def\eea{\end{eqnarray}}
\def\ba{\begin{array}}
\def\ea{\end{array}}
\def\no{\nonumber}
\def\le{\langle}
\def\re{\rangle}
\def\lt{\left}
\def\rt{\right}

\newtheorem{Theorem}{Theorem}
\newtheorem{Definition}{Definition}
\newtheorem{Proposition}{Proposition}
\newtheorem{Lemma}{Lemma}
\newtheorem{Corollary}{Corollary}
\newcommand{\proof}[1]{{\bf Proof. }
        #1\begin{flushright}$\Box$\end{flushright}}

\baselineskip=20pt

\newfont{\elevenmib}{cmmib10 scaled\magstep1}
\newcommand{\preprint}{
   \begin{flushright}\normalsize  \sf
     {\tt hep-th/0411160} \\ October 2004
   \end{flushright}}
\newcommand{\Title}[1]{{\baselineskip=26pt
   \begin{center} \Large \bf #1 \\ \ \\ \end{center}}}
\newcommand{\Author}{\begin{center}
   \large \bf
Wen-Li Yang${}^{a,b}$
 ~ and~Yao-Zhong Zhang${}^b$\end{center}}
\newcommand{\Address}{\begin{center}

     ${}^a$ Institute of Modern Physics, Northwest University
     Xian 710069, P.R. China\\
     ~~\\
     ${}^b$ Department of Mathematics, The University of Queensland,
     Brisbane 4072, Australia
   \end{center}}
\newcommand{\Accepted}[1]{\begin{center}
   {\large \sf #1}\\ \vspace{1mm}{\small \sf Accepted for Publication}
   \end{center}}

\preprint
\thispagestyle{empty}
\bigskip\bigskip\bigskip

\Title{Non-diagonal solutions of the  reflection equation for the
trigonometric $A^{(1)}_{n-1}$ vertex  model } \Author

\Address
\vspace{1cm}

\begin{abstract}
We obtain a class of non-diagonal solutions of the  reflection
equation for the  trigonometric $A^{(1)}_{n-1}$ vertex model. The
solutions can be expressed in terms of intertwinner matrix and its
inverse, which intertwine two trigonometric R-matrices. In
addition to  a {\it discrete\/} (positive integer) parameter $l$,
$1\leq l\leq n$, the solution contains $n+2$ {\it continuous\/}
boundary parameters.

\vspace{1truecm}

\end{abstract}
\newpage
\section{Introduction}
\label{intro} \setcounter{equation}{0}

Two-dimensional integrable models have traditionally been solved
by imposing periodic boundary conditions. For such bulk systems,
the quantum Yang-Baxter equation (QYBE)\bea
R_{12}(u_1-u_2)R_{13}(u_1-u_3)R_{23}(u_2-u_3)=
R_{23}(u_2-u_3)R_{13}(u_1-u_3)R_{12}(u_1-u_2), \label{QYB} \eea
leads to families of commuting row-to-row {\it transfer matrices}
which may be diagonalized  by the quantum inverse scattering
method (QISM) \cite{Kor93}. The QYBE has been studied for a long
time and a large number of solutions, the so-called R-matrices,
are known.

Not all boundary conditions are compatible with integrability in
the bulk. The bulk integrability only holds when one imposes
integrable boundary conditions specified by the so-called boundary
K-matrices $K^{\pm}(u)$ which satisfy the reflection equation (RE)
and its dual \cite{Skl88}-\cite{Gho94}. The RE (also called the
boundary Yang-Baxter equation) and its dual are the analogue of
the QYBE  for  models with open boundaries.

Much effort has been made on constructing ({\it non-diagonal\/})
solutions (K-matrices)  of the RE by brute force (i.e. solving the
RE directly) \cite{Gho94}-\cite{Yan041}, and more recently by the
boundary quantum group approach \cite{Mez98}-\cite{Arn03} and the
affine Heck algebra approach \cite{Doi04}. In this paper, we
propose a different approach, which is referred to as the
intertwiner-matrix approach, to construct a class of non-diagonal
solutions of the RE associated with the trigonometric
$A^{(1)}_{n-1}$ vertex model. The idea is originated from the
studies \cite{Fan98,Yan041} of the RE associated with the $\Zb_n$
elliptic Belavin model \cite{Bel81}. It has been observed in
\cite{Yan043} that the generic non-diagonal solutions
\cite{Gho94,Dev93} of the RE associated with the trigonometric
$A^{(1)}_{1}$ vertex model (spin-$\frac{1}{2}$ XXZ model), as in
the $\Zb_n$ Belavin model, can be expressed in terms of the
intertwiner-matrices  and the diagonal face-type K-matrix. The
advantage of this face-vertex correspondence method is that in the
``face" picture the corresponding K-matrices become diagonal. This
enables one to diagonalize the corresponding double-row transfer
matrix with the special choice of the left and right boundary
parameters \cite{Yan042} by the generalized QISM developed in
\cite{Yan04}. Despite the success of this approach in the
trigonometric $A^{(1)}_1$ and $\Zb_n$ elliptic vertex boundary
models, to our knowledge the face-vertex correspondence for
boundary trigonometric $A^{(1)}_{n-1}$ model with $n>2$ has not
been achieved so far.  This is also in contrast to the boundary
rational $A^{(1)}_{n-1}$ model where the non-diagonal K-matrices
are equivalent to  the diagonal ones by simple
spectral-independent similarity transformations
\cite{Min01,Gal04}.

We have organized the paper as follows. In section 2, we introduce
our notation and some  basic ingredients. In section 3, we
construct the intertwiner-matrix which satisfies the face-vertex
correspondence relation between the two R-matrices $R(u)$ and
$W(u)$. Through the {\it magic\/} intertwiner-matrix, we obtain,
in section 4, a set of non-diagonal solutions to the RE and the
dual RE. Section 5 is devoted to the conclusion. A sketch of the
proof of the face-vertex correspondence relation is provided  in
the Appendix.

\section{$A^{(1)}_{n-1}$ reflection equation}
 \label{RE} \setcounter{equation}{0}

Let us fix a positive integer $n$ ($n\geq 2$) and a generic
complex number $\eta$, and  $R(u)\in End(\Cb^n\otimes\Cb^n)$ be
the trigonometric solution to the $A^{(1)}_{n-1}$ type QYBE given
by \cite{Che80,Per81,Baz91} \bea
\hspace{-0.5cm}R(u)=\sum_{\a=1}^{n}R^{\a\a}_{\a\a}(u)E_{\a\a}\otimes
E_{\a\a} +\sum_{\a\ne \b}\lt\{R^{\a\b}_{\a\b}(u)E_{\a\a}\otimes
E_{\b\b}+ R^{\b\a}_{\a\b}(u)E_{\b\a}\otimes
E_{\a\b}\rt\},\label{R-matrix} \eea where $E_{ij}$ is the matrix
with elements $(E_{ij})^l_k=\d_{jk}\d_{il}$. The coefficient
functions are \bea
R^{\a\b}_{\a\b}(u)&=&\lt\{\begin{array}{cc}\frac{\sin(u)\,e^{-i\eta}}{\sin(u+\eta)},&
\a>\b,\\[6pt]1,&\a=\b,\\[6pt]\frac{\sin(u)\,e^{i\eta}}{\sin(u+\eta)},&
\a<\b,\end{array}\rt.,\label{Elements1}\\[6pt]
R^{\b\a}_{\a\b}(u)&=&\lt\{\begin{array}{cc}\frac{\sin(\eta)\,e^{iu}}{\sin(u+\eta)},&
\a>\b,\\[6pt]1,&\a=\b,\\[6pt]\frac{\sin(\eta)\,e^{-iu}}{\sin(u+\eta)},&
\a<\b,\end{array}\rt..\label{Elements2}\eea One can check that the
R-matrix satisfies the following unitarity, crossing-unitarity and
quasi-classical relations:\begin{eqnarray}
 &&\hspace{-1.5cm}\mbox{
 Unitarity}:\hspace{42.5mm}R_{12}(u)R_{21}(-u)= {\rm id},\label{Unitarity}\\
 &&\hspace{-1.5cm}\mbox{
 Crossing-unitarity}:\quad
 R^{t_2}_{12}(u)M_2^{-1}R_{21}^{t_2}(-u-n\eta)M_2
 = \frac{\sin(u)\sin(u+n\eta)}{\sin(u+\eta)\sin(u+n\eta-\eta)}\,\mbox{id},
 \label{crosing-unitarity}\\
 &&\hspace{-1.5cm}\mbox{ Quasi-classical
 property}:\hspace{22.5mm}\, R_{12}(u)|_{\eta\rightarrow 0}= {\rm
id}.\label{quasi}
\end{eqnarray}
Here $R_{21}(u)=P_{12}R_{12}(u)P_{12}$ with $P_{12}$ being the
usual permutation operator and $t_i$ denotes the transposition in
the $i$-th space, and $\eta$ is the so-called crossing paramter.
The crossing matrix $M$ is a diagonal $n\times n$ matrix with
elements \bea
M_{\a\b}=M_{\a}\d_{\a\b},~~M_{\a}=e^{-2i\a\eta},~\a=1,\cdots,n.\label{C-Matrix}\eea
Here and below we adopt the standard notation: for any matrix
$A\in {\rm End}(\Cb^n)$, $A_j$ is an embedding operator in the
tensor space $\Cb^n\otimes \Cb^n\otimes\cdots$, which acts as $A$
on the $j$-th space and as an identity on the other factor spaces;
$R_{ij}(u)$ is an embedding operator of R-matrix in the tensor
space, which acts as an identity on the factor spaces except for
the $i$-th and $j$-th ones.

One introduces  the ``row-to-row" monodromy matrix $T(u)$, which
is an $n\times n$ matrix with elements being operators acting  on
$(\Cb^n)^{\otimes N}$  \begin{eqnarray}
T(u)=R_{01}(u+z_1)R_{02}(u+z_2)\cdots
R_{0N}(u+z_N).\label{T-matrix}\end{eqnarray} Here
$\{z_i|i=1,\cdots, N\}$ are arbitrary free complex parameters
which are usually called inhomogeneous parameters. With the help
of the QYBE (\ref{QYB}), one can show that $T(u)$ satisfies the
so-called ``RLL" relation
\begin{eqnarray}
R_{12}(u-v)T_1(u)T_2(v)=T_2(v)T_1(u)R_{12}(u-v).\label{Relation1}\end{eqnarray}

Integrable open chains can be constructed as follows \cite{Skl88}.
Let us introduce the K-matrix $K^-(u)$ which gives rise to an
integrable boundary condition on the right boundary. $K^-(u)$
satisfies  the RE
 \begin{eqnarray}
 &&R_{12}(u_1-u_2)K^-_1(u_1)R_{21}(u_1+u_2)K^-_2(u_2)\no\\
  &&~~~~~~=
 K^-_2(u_2)R_{12}(u_1+u_2)K^-_1(u_1)R_{21}(u_1-u_2).\label{RE-V}
\end{eqnarray}
For models with open boundaries, instead of the standard
``row-to-row" monodromy matrix $T(u)$ (\ref{T-matrix}), one needs
the  ``double-row" monodromy matrix $\mathcal{T}(u)$
\begin{eqnarray}
 \mathcal{T}(u)=T(u)K^-(u)T^{-1}(-u).\label{Mon-V-1}
\end{eqnarray}
Using (\ref{Relation1}) and (\ref{RE-V}), one can prove that
$\mathcal{T}(u)$ satisfies
\begin{eqnarray}
 R_{12}(u_1-u_2)\mathcal{T}_1(u_1)R_{21}(u_1+u_2)
  \mathcal{T}_2(u_2)=
 \mathcal{T}_2(u_2)R_{12}(u_1+u_2)\mathcal{T}_1(u_1)R_{21}(u_1-u_2).
 \label{Relation-Re}
\end{eqnarray}
In order to construct the {\it double-row transfer matrices},
besides the RE, one  needs another K-matrix $K^+(u)$ which gives
integrable boundary condition on the left boundary. The explicit
form of the dual RE is related with the crossing-unitarity
relation of the R-matrix \cite{Skl88,Mez91}. For a R-matrix whose
partial transpositions have inverse, the corresponding dual RE
reads \bea
&&R_{12}(u_2-u_1)K^+_1(u_1)R_{21}^{-1,t_2,-1,t_2}(-u_1-u_2)K^+_2(u_2)\no\\
&&\qquad\quad=
K^+_2(u_2)R_{12}^{-1,t_1,-1,t_1}(-u_1-u_2)K^{+}_1(u_1)R_{21}(u_2-u_1).
\label{DRE-V}\eea The crossing-unitarity relation
(\ref{crosing-unitarity}) of the R-matrix $R(u)$ allows one to
simplify the dual RE into the form \bea
&&R_{12}(u_2-u_1)K^+_1(u_1)\,M_1^{-1}\,R_{21}(-u_1-u_2-n\eta)\,M_1\,K^+_2(u_2)\no\\
&&\qquad\quad=
M_1\,K^+_2(u_2)R_{12}(-u_1-u_2-n\eta)\,M_1^{-1}\,K^{+}_1(u_1)R_{21}(u_2-u_1).
\label{DRE-V1}\eea Different  integrable boundary conditions are
described by different solutions $K^{-}(u)$ ($K^{+}(u)$) to the
(dual) RE \cite{Skl88, Gho94}. Then the {\it double-row transfer
matrix\/} of  the inhomogeneous model associated with the R-matrix
(\ref{R-matrix})-(\ref{Elements2}) with open boundary specified by
the K-matrices $K^{\pm}(u)$ is given by
\begin{eqnarray}
\tau(u)=tr(K^+(u)\mathcal{T}(u)).\label{trans}
\end{eqnarray} The commutativity of the transfer matrices
\begin{eqnarray}
 [\tau(u),\tau(v)]=0,\label{Com-2}
\end{eqnarray}
follows as a consequence of (\ref{QYB}),
(\ref{Unitarity})-(\ref{crosing-unitarity}) and
(\ref{Relation-Re})-(\ref{DRE-V}). This ensures the integrability
of the inhomogeneous  model with open boundary. In this paper, we
search for non-diagonal solutions to the RE (\ref{RE-V}) with
multiple continuous parameters (c.f. \cite{Dev93}) by means of the
intertwiner-matrix approach.

\section{Intertwining vectors and the associated face-vertex correspondence relations}
 \label{IRF} \setcounter{equation}{0}

Let $\lt\{\e_{i}~|~i=1,2,\cdots,n\rt\}$ be the orthonormal basis
of the vector space $\Cb^n$ such that $\langle\e_i,~\e_j
\rangle=\d_{ij}$. For a vector $m\in \Cb^n$, define \bea
m_i=\langle m,\e_i\rangle, ~~|m|=\sum_{l=1}^nm_l,~~i=1,\cdots,n.
\label{Def1}\eea Let us introduce $n$ intertwiners
$\{\phi_{m,m-\e_j}(u)|\, j=1,\cdots,n\}$. Each
$\phi_{m,m-\e_j}(u)$ is an $n$-component column vector whose
$\a$-th elements are $\{\phi^{(\a)}_{m,m-\e_j}(u)\}$. The $n$
intertwiners form an $n\times n$ matrix (in which $j$ and $\a$
stand for the column and the row indices respectively), called the
intertwiner-matrix,  with the non-vanishing matrix elements being
\bea \hspace{-0.4truecm}\lt(\begin{array}{ccccccc}e^{i\eta f_1(m)}
&&&&&&e^{i\eta
F_n(m)+\rho_n}e^{2iu}\\e^{i\eta F_1(m)+\rho_1}&e^{i\eta f_2(m)}&&&&&\\
&e^{i\eta F_2(m)+\rho_2}&\ddots&&&&\\&&\ddots&e^{i\eta
f_j(m)}&&&\\&&&e^{i\eta
F_j(m)+\rho_j}&\ddots&&\\&&&&\ddots&e^{i\eta
f_{n-1}(m)}&\\&&&&&e^{i\eta F_{n-1}(m)+\rho_{n-1}}&e^{i\eta
f_n(m)}
\end{array}\rt).\label{In-matrix-0}\eea Here $\{\rho_i|i=1,\cdots,n\}$ are complex
constants with regard to $u$ and $m$,  and
$\{f_i(m)|i=1,\cdots,n\}$ and $\{F_{i}(m)|i=1,\cdots,n\}$ are
linear functions of $m$:\bea
f_i(m)&=&\sum_{l=1}^{i-1}m_l-m_i-\frac{1}{2}|m|,~~i=1,\cdots,n,\label{function1}\\
F_{i}(m)&=&\sum_{l=1}^{i}m_l-\frac{1}{2}|m|,~~i=1,\cdots,n-1,\label{function2}\\
F_n(m)&=&-\frac{3}{2}|m|.\label{function3}\eea

By the definitions of  $f_j(m)$ and $F_j(m)$, after a
straightforward calculation, we find \bea
f_j(\e_k)&=&\lt\{\begin{array}{cc}\frac{1}{2},&j>k,\\
[4pt]-\frac{3}{2},&j=k,\\[4pt]-\frac{1}{2},&j<k,
\end{array}\rt.\label{f-fuction-1}\\[6pt]
F_j(\e_k)&=&\lt\{\begin{array}{cc}\frac{1}{2},&j\geq k,\\
[4pt]-\frac{1}{2},&j<k,
\end{array}\rt.~~{\rm for}~j\neq n,\label{f-fuction-2}\\
[4pt]F_n(\e_k)&=&-\frac{3}{2}. \label{f-fuction-3}\eea The above
equations allow us to derive the following {\it face-vertex
correspondence relation\/}:\bea &&R_{12}(u_1-u_2)
\phi_{m,m-\e_i}(u_1)\otimes
\phi_{m-\e_i,m-\e_i-\e_j}(u_2) \no\\
&&~~~~=\sum_{kl}W^{kl}_{ij}(u_1-u_2)
\phi_{m-\e_{l},m-\e_{l}-\e_{k}}(u_1)\otimes
\phi_{m,m-\e_{l}}(u_2).\label{Face-vertex}\eea Here the
non-vanishing elements of $\{W(u)^{kl}_{ij}\}$ are \bea
W^{jj}_{jj}(u)&=&1,~~W^{jk}_{jk}(u)=\frac{\sin(u)}{\sin(u+\eta)},
~~{\rm for}\,\,j\neq k,\label{W-elements-1}\\[6pt]
W^{kj}_{jk}(u)&=&\lt\{\begin{array}{cc}\frac{\sin(\eta)\,e^{iu}}{\sin(u+\eta)},&
j>k,\\[6pt]\frac{\sin(\eta)\,e^{-iu}}{\sin(u+\eta)},&
j<k,\end{array}\rt.~~{\rm for}~j\neq k.\label{W-elements-2}\eea A
sketch of proof of the above face-vertex correspondence relation
is relegated to the Appendix. Then associated with
$\{W(u)^{kl}_{ij}\}$, one can introduce ``{\it face type\/}"
 R-matrix $W(u)$ \bea
W(u)=\sum_{i,j,k,l} W^{kl}_{ij}(u)E_{ki}\otimes
E_{lj}.\label{W-matrix}\eea Some remarks are in order.
 The face type R-matrix $W(u)$ does not depend on the face type parameter $m$, in contrast to
 the $\Zb_n$ elliptic case \cite{Bel81,Jim87}.
 It follows that $W(u)$ and $R(u)$
 satisfy the same QYBE, i.e. $W(u)$ obeys the usual (vertex type) QYBE rather than
 the {\it dynamical\/} one \cite{Hou03,Yan04}.

The forms of the R-matrices $R(u)$
(\ref{R-matrix})-(\ref{Elements2}) and $W(u)$
(\ref{W-elements-1})-(\ref{W-matrix}) ensure that the set of
intertwiner matrices satisfying the face-vertex correspondence
relation (\ref{Face-vertex}) is invariant under the similarity
transformation given by  an arbitrary {\it diagonal constant\/}
matrix. This allows one to turn the intertwiner-matrix
(\ref{In-matrix-0}) into the following  {\it standard\/} form:
\bea \lt(\begin{array}{ccccccc}e^{i\eta f_1(m)} &&&&&&e^{i\eta
F_n(m)+\rho}e^{2iu}\\e^{i\eta F_1(m)}&e^{i\eta f_2(m)}&&&&&\\
&e^{i\eta F_2(m)}&\ddots&&&&\\&&\ddots&e^{i\eta
f_j(m)}&&&\\&&&e^{i\eta F_j(m)}&\ddots&&\\&&&&\ddots&e^{i\eta
f_{n-1}(m)}&\\&&&&&e^{i\eta F_{n-1}(m)}&e^{i\eta f_n(m)}
\end{array}\rt),\label{In-matrix}\eea  where $\rho$ is related to the
original parameters $\{\rho_i\}$ in (\ref{In-matrix-0}) by
$\rho=\sum_{l=1}^n\rho_l$, and the linear functions $\{f_j(m)\}$
and $\{F_j(m)\}$ are the same as those given by
(\ref{function1})-(\ref{function3}). Noting that \bea
\sum_{i=1}^nf_i(m)=\sum_{i=1}^nF_i(m)=\sum_{l=1}^{n}\frac{n-2(l+1)}{2}m_l,\eea
one can show that the determinant of the intertwiner matrix
(\ref{In-matrix}) is \bea {\rm
Det}\lt(\phi^{(\a)}_{m,m-\e_j}(u)\rt)=e^{i\eta\sum_{l=1}^{n}\frac{n-2(l+1)}{2}m_l}\,
(1-(-1)^ne^{2iu+\rho}).\label{Det}\eea For a generic $\rho\in\Cb$
this determinant is not vanishing and thus the inverse of
(\ref{In-matrix}) exists. This fact allows us to introduce other
types of intertwiners $\bar{\phi}$ and $\tilde{\phi}$ satisfying
the following orthogonality conditions: \bea
&&\sum_{\a}\bar{\phi}^{(\a)}_{m,m-\e_i}(u)
~\phi^{(\a)}_{m,m-\e_j}(u)=\d_{ij},\label{Int1}\\[6pt]
&&\sum_{\a}\tilde{\phi}^{(\a)}_{m+\e_i,m}(u)
~\phi^{(\a)}_{m+\e_j,m}(u)=\d_{ij}.\label{Int2}\eea From these
conditions we  derive the ``completeness" relations:\bea
&&\sum_{k}\bar{\phi}^{(\a)}_{m,m-\e_k}(u)
~\phi^{(\b)}_{m,m-\e_k}(u)=\d_{\a\b},\label{Int3}\\[6pt]
&&\sum_{k}\tilde{\phi}^{(\a)}_{m+\e_{k},m}(u)
~\phi^{(\b)}_{m+\e_{k},m}(u)=\d_{\a\b}.\label{Int4}\eea

With the help of (\ref{Int1})-(\ref{Int4}), we  obtain, from the
face-vertex correspondence relation (\ref{Face-vertex}),
\begin{eqnarray}
 &&\left(\tilde{\phi}_{m+\e_{k},m}(u_1)\otimes
 {\rm id}\right)R_{12}(u_1-u_2) \left({\rm
id}\otimes\phi_{m+\e_{j},m}(u_2)\right)\no\\
 &&\qquad\quad= \sum_{i,l}W^{kl}_{ij}(u_1-u_2)\,
 \tilde{\phi}_{m+\e_{i}+\e_{j},m+\e_{j}}(u_1)\otimes
 \phi_{m+\e_{k}+\e_{l},m+\e_{k}}(u_2),\label{Face-vertex1}\\
 &&\left(\tilde{\phi}_{m+\e_{k},m}(u_1)\otimes
 \tilde{\phi}_{m+\e_{k}+\e_{l},m+\e_{k}}(u_2)\right)R_{12}(u_1-u_2)\no\\
 &&\qquad\quad= \sum_{i,j}W^{kl}_{ij}(u_1-u_2)\,
 \tilde{\phi}_{m+\e_{i}+\e_{j},m+\e_{j}}(u_1)\otimes
 \tilde{\phi}_{m+\e_{j},m}(u_2),\label{Face-vertex2}\\
 &&\left({\rm id}\otimes
 \bar{\phi}_{m,m-\e_{l}}(u_2)\right)R_{12}(u_1-u_2)
 \left(\phi_{m,m-\e_{i}}(u_1)\otimes {\rm id}\right)\no\\
 &&\qquad\quad= \sum_{k,j}W^{kl}_{ij}(u_1-u_2)\,
 \phi_{m-\e_{l},m-\e_{k}-\e_{l}}(u_1)\otimes
 \bar{\phi}_{m-\e_{i},m-\e_{i}-\e_{j}}(u_2),\label{Face-vertex3}\\
 &&\left(\bar{\phi}_{m-\e_{l},m-\e_{k}-\e_{l}}(u_1)\otimes
 \bar{\phi}_{m,m-\e_{l}}(u_2)\right)R_{12}(u_1-u_2)\no\\
 &&\qquad\quad= \sum_{i,j}W^{kl}_{ij}(u_1-u_2)\,
 \bar{\phi}_{m,m-\e_{i}}(u_1)\otimes
 \bar{\phi}_{m-\e_{i},m-\e_{i}
 -\e_{j}}(u_2).\label{Face-vertex4}
\end{eqnarray}

\section{Non-diagonal solutions to the reflection equation}
\label{NRE} \setcounter{equation}{0}

We first present the main new result of this paper. It can be
shown (see subsection 4.1) that  \bea K^-(u)^s_t=\sum_{i=1}^n
k_i(u)\phi^{(s)}_{\l,\l-\e_{i}}(u)
\bar{\phi}^{(t)}_{\l,\l-\e_{i}}(-u), \label{K-matrix}\eea gives a
 class of non-diagonal solutions
of the RE associated with the trigonometric R-matrix
(\ref{R-matrix})-(\ref{Elements2}). Here
 $\{k_i(u)|i=1,\cdots,n\}$ are the matrix elements of the
 $\l$-independent face-type diagonal matrix $\K(\l|u)$,
 \bea  \K(\l|u)={\rm
Diag\/}\lt(k_1(u),\cdots,k_n(u)\rt),\label{D-K-1}\eea where
\bea k_j(u)=\lt\{\begin{array}{ll}1,&1\leq j\leq l,\\[4pt]
\frac{\sin(\xi-u)}{\sin(\xi+u)}e^{-2iu},&l+1\leq j\leq
n.\end{array}\rt.\label{K-matrix1}\eea $\phi_{\l,\l-\e_{i}}(u)$
and $\bar{\phi}_{\l,\l-\e_{i}}(u)$  are respectively given by the
$i$-th column of (\ref{In-matrix}) and (\ref{Int1}) specializing
to $m=\l=\sum_{i=1}^n\l_i\e_i$, and  $l$ is  an positive integer,
$1\leq l\leq n$. In addition to the discrete boundary parameter
$l$ (c.f. \cite{Dev93}), the solution contains $n+2$ {\it
continuous\/} boundary parameters $\xi$, $\rho$ and
$\{\l_i|i=1,\cdots,n\}$. Besides the RE (\ref{RE-V}) (which will
be proved later), the K-matrix  $K^-(u)$ (\ref{K-matrix}) also
satisfies the {\it regular\/} condition (c.f.
\cite{Gan99,Del01}):\bea K^-(0)={\rm id},\label{Reg}\eea and
boundary unitarity relation:\bea K^-(u)\,K^-(-u)={\rm
id}.\label{b-Uni}\eea (\ref{Reg}) and (\ref{b-Uni}) follow from
the orthogonality condition (\ref{Int1}) and the  ``completeness"
relation (\ref{Int3}) specializing to $m=\l$. In the following, we
shall show that the non-diagonal K-matrix  $K^-(u)$
(\ref{K-matrix}) satisfies the RE (\ref{RE-V}) with the R-matrix
given by (\ref{R-matrix})-(\ref{Elements2}).

\subsection{Proof of the RE}
Let $K^-(u)$ be a solution to the RE (\ref{RE-V}). As in
\cite{Yan041}, we introduce the corresponding  face-type K-matrix
$\K(m|u)$ via \bea &&\K(m|u)^j_i=\sum_{s,t}\tilde{\phi}^{(s)}
_{m-\e_{i}+\e_{j},~\l-\e_{i}}(u)K^-(u)^s_t\phi^{(t)}
_{m,~m-\e_{i}}(-u).\label{F-V1}\eea

\noindent Multiplying both sides of the RE (\ref{RE-V}) from the
right by $\phi_{m+\e_{i_3},\,m}(-u_1) \otimes
\phi_{m+\e_{i_3}+\e_{j_3},\,m+\e_{i_3}}(-u_2)$, and using the
face-vertex correspondence relation (\ref{Face-vertex}) and the
``completeness" relation (\ref{Int4}), we have, for the L.H.S. of
the resulting relation, \bea
{\rm L.H.S.}&=&R_{12}(u_1-u_2)K^-_1(u_1)R_{21}(u_1+u_2)\no\\
&&\qquad\quad\times (\phi_{m+\e_{i_3},\,m}(-u_1)\otimes K^-(u_2)\,
\phi_{m+\e_{i_3}+\e_{j_3},\,m+\e_{i_3}}(-u_2))\no\\
&=&R_{12}(u_1-u_2)K^-_1(u_1)R_{21}(u_1+u_2)
(\phi_{m+\e_{i_3},\,m}(-u_1)\otimes 1)\no\\
&&\qquad\quad\times(1\otimes \{\sum_{j_2}
\phi_{m+\e_{i_3}+\e_{j_2},\,m+\e_{i_3}}(u_2)
\tilde{\phi}_{m+\e_{i_3}+\e_{j_2},\,m+\e_{i_3}}(u_2)\no\\
&&\qquad\quad\times \lt.K^-(u_2)
\phi_{m+\e_{i_3}+\e_{j_3},\,m+\e_{i_3}}(-u_2)\rt\})\no\\
&=&\sum_{j_2}R_{12}(u_1-u_2)K^-_1(u_1)R_{21}(u_1+u_2)\no\\
&&\qquad\quad\times\lt(\phi_{m+\e_{i_3},\,m}(-u_1)\otimes
\phi_{m+\e_{i_3}+\e_{j_2},\,m+\e_{i_3}}(u_2)\rt)
\K(m+\e_{i_3}+\e_{j_3}|u_2)^{j_2}_{j_3}
\no\\
&=&\sum_{i_2}\sum_{j_1,j_2}R_{12}(u_1-u_2)K^-_1(u_1)
W^{j_1\,i_2}_{j_2\,i_3}(u_1+u_2)\no\\
&&\qquad\quad\times
\lt(\phi_{m+\e_{i_3}+\e_{j_2},\,m+\e_{j_1}}(-u_1)\otimes
\phi_{m+\e_{j_1},\,m}(u_2)\rt)\K(m+\e_{i_3}+\e_{j_3}|u_2)^{j_2}_{j_3}
\no\\
&&\vdots\no\\
&=&\sum_{i_0,j_0} (\phi_{m+\e_{i_0},\,m}(u_1)\otimes
\phi_{m+\e_{i_0}+\e_{j_0},\,m+\e_{i_0}}(u_2))\no\\
&&\qquad\quad\times \lt\{\sum_{i_1,i_2}
\sum_{j_1,j_2}W^{i_0\,j_0}_{i_1\,j_1}(u_1-u_2)
\K(m+\e_{i_2}+\e_{j_1}|u_1)^{i_1}_{i_2}\rt.\no\\
&&\qquad\quad\qquad\quad\times
\lt.W^{j_1\,i_2}_{j_2\,i_3}(u_1+u_2)
\K(m+\e_{i_3}+\e_{j_3}|u_2)^{j_2}_{j_3}\rt\}.\label{LHS} \eea
Similarly for the R.H.S. of the resulting relation, we obtain \bea
{\rm R.H.S.}&=& \sum_{i_0,j_0}
\lt(\phi_{m+\e_{i_0},\,m}(u_1)\otimes
\phi_{m+\e_{i_0}+\e_{j_0},\,m+\e_{i_0}}(u_2)\rt)\no\\
&&\qquad\quad\times\lt\{\sum_{i_1,i_2}
\sum_{j_1,j_2}\K(\l+\e_{i_0}+\e_{j_1}|u_2)^{j_0}_{j_1}
W^{i_0\,j_1}_{i_1\,j_2}(u_1+u_2)\rt.
\no\\
&&\qquad\quad\qquad\quad\times
\lt.\K(m+\e_{i_2}+\e_{j_2}|u_1)^{i_1}_{i_2}
W^{j_2\,i_2}_{j_3\,i_3}(u_1-u_2)\rt\}.\label{RHS} \eea Note that
intertwiners are linearly independent, which follows from
(\ref{Det}). Therefore the K-matrix $K^-(u)$ would satisfy the RE
(\ref{RE-V}) provided that the corresponding face-type K-matrix
$\K(m|u)$ defined by (\ref{F-V1}) obeys the relation
 \bea
&&\sum_{i_1,i_2}\sum_{j_1,j_2}~
W^{i_0\,j_0}_{i_1\,j_1}(u_1-u_2)\K(m+\e_{j_1}+\e_{i_2}|u_1)
^{i_1}_{i_2}\no\\
&&\qquad\quad\qquad\quad\times W^{j_1\,i_2}_{j_2\,i_3}(u_1+u_2)
\K(m+\e_{j_3}+\e_{i_3}|u_2)^{j_2}_{j_3}\no\\
&&~=\sum_{i_1,i_2}\sum_{j_1,j_2}~ \K(m+\e_{j_1}+\e_{i_0}|u_2)
^{j_0}_{j_1}W^{i_0\,j_1}_{i_1\,j_2}(u_1+u_2)\no\\
&&\qquad\quad\qquad\quad\times
\K(m+\e_{j_2}+\e_{i_2}|u_1)^{i_1}_{i_2}
W^{j_2\,i_2}_{j_3\,i_3}(u_1-u_2).\label{RE-F} \eea Specializing
$m$ to the boundary parameter $\l$, i.e. $m=\l$, then one can
easily check that the $\l$-independent face-type diagonal K-matrix
$ \K(\l|u)$ given by (\ref{D-K-1})-(\ref{K-matrix1}) solves
(\ref{RE-F}). Therefore, the non-diagonal K-matrix $K^-(u)$ given
by (\ref{K-matrix})-(\ref{K-matrix1}) satisfies the RE
(\ref{RE-V}) associated with the R-matrix $R(u)$ given by
(\ref{R-matrix})-(\ref{Elements2}).

\subsection{Solution of the dual RE}

Similarly, one can show  that the non-diagonal K-matrix $K^+(u)$
given by  \bea K^+(u)^s_t=\sum_{i=1}^n
k^+_i(u)\phi^{(s)}_{\l',\l'-\e_{i}}(-u)
\tilde{\phi}^{(t)}_{\l',\l'-\e_{i}}(u), \label{DK-matrix}\eea
where
\bea k^+_j(u)=\lt\{\begin{array}{ll}e^{-2i(j\eta)},&1\leq j\leq l',\\[4pt]
\frac{\sin(\bar{\xi}+u+\frac{n}{2}\eta)}{\sin(\bar{\xi}-u-\frac{n}{2}\eta)}
e^{2i(u+\frac{n-2j}{2}\eta)},&l'+1\leq j\leq
n,\end{array}\rt.\label{DK-matrix1}\eea satisfies the dual RE
(\ref{DRE-V1}). Here $\phi_{\l',\l'-\e_{i}}(u)$ and
$\tilde{\phi}_{\l',\l'-\e_{i}}(u)$ are respectively given by the
$i$-th column of (\ref{In-matrix}) (with $\rho$ replaced by
$\rho'$) and (\ref{Int2}) specializing to $m=\l'$, and  $l'$ is an
positive integer, $1\leq l'\leq n$, (which  may be different from
$l$). In addition to the discrete boundary parameter $l'$, the
solution contains $n+2$ {\it continuous\/} boundary parameters
$\bar{\xi}$, $\rho'$ and $\{\l'_i|i=1,\cdots,n\}$ (which  may be
different from $\xi$, $\rho$ and $\{\l_i|i=1,\cdots,n\}$,
respectively).

\section{Discussions}
\label{Con} \setcounter{equation}{0}

We have constructed the intertwiner-matrix $\phi$
(\ref{In-matrix}) between the two trigonometric $A^{(1)}_{n-1}$
R-matrices  $R(u)$, (\ref{R-matrix})-(\ref{Elements2}), and
$W(u)$, (\ref{W-elements-1})-(\ref{W-matrix}). From the
intertwiner-matrix and its associated matrices $\bar{\phi}$
(\ref{Int1}) and $\tilde{\phi}$ (\ref{Int2}), we obtain a class of
nondiagonal solutions $K^-(u)$, given by
(\ref{K-matrix})-(\ref{K-matrix1}), of the RE (resp. a class of
nondiagonal solutions $K^{+}(u)$, given by
(\ref{DK-matrix})-(\ref{DK-matrix1}), of the dual RE) associated
with the R-matrix $R(u)$. In addition to the discrete parameter
$l$ (resp. $l'$), the solution contains $n+2$ continuous boundary
parameters $\xi$, $\rho$ and $\{\l_i|i=1,\cdots,n\}$ (resp.
$\bar{\xi}$, $\rho'$ and $\{\l'_i|i=1,\cdots,n\}$).

Consider the special cases where some of the two sets of boundary
parameters associated with the left and right boundaries are
frozen by the restrictions,\bea \rho'=\rho,~~
\l'+\sum_{l=1}^N\e_{i_l}=\l,\label{Restriction}\eea where $N$ is a
positive integer, $\{i_l|l=1,\cdots,N\}$ are positive integers
such that $ 2\leq i_l\leq n$. Then the decompositions of the
K-matrices $K^{\pm}(u)$ (\ref{K-matrix}) and (\ref{DK-matrix}) in
terms of the intertwiner-matrices and diagonal face-type
K-matrices should enable us to diagonalize the double-row transfer
matrix of the trigonometric $A^{(1)}_{n-1}$ vertex model  with
non-diagonal open boundaries  by means of  the generalized
algebraic Bethe ansatz method developed in \cite{Yan04}. The
results will be presented elsewhere \cite{Yan042}.

\section*{Acknowledgements}
This work was financially supported by the Australian Research
Council.

\section*{Appendix: The face-vertex correspondence relation}
\setcounter{equation}{0}

\renewcommand{\theequation}{A.\arabic{equation}}
Thanks to  the non-vanishing matrix elements of the R-matrices
$R(u)$ (\ref{R-matrix})-(\ref{Elements2}) and $W(u)$
(\ref{W-elements-1})-(\ref{W-matrix}), the proof of the
face-vertex correspondence relation is  reduced to the proof of
the relations:\bea &&R_{12}(u_1-u_2) \phi_{m,m-\e_i}(u_1)\otimes
\phi_{m-\e_i,m-2\e_i}(u_2) \no\\
&&\quad=W^{ii}_{ii}(u_1-u_2) \phi_{m-\e_{i},m-2\e_{i}}(u_1)\otimes
\phi_{m,m-\e_{i}}(u_2),\label{E-1}\\[6pt]
&&\phi^{(\a)}_{m,m-\e_i}(u_1)\,
\phi^{(\a)}_{m-\e_i,m-\e_i-\e_j}(u_2)=W^{ij}_{ij}(u_1-u_2)
\phi^{(\a)}_{m-\e_{j},m-\e_{j}-\e_i}(u_1)\,
\phi^{(\a)}_{m,m-\e_{j}}(u_2)\no\\
&&\qquad\quad+W^{ji}_{ij}(u_1-u_2)
\phi^{(\a)}_{m-\e_{i},m-\e_{i}-\e_j}(u_1)\,
\phi^{(\a)}_{m,m-\e_{i}}(u_2),~~i\neq j,\label{E-2}\\[6pt]
&&R^{\a\b}_{\a\b}(u_1-u_2) \phi^{(\a)}_{m,m-\e_i}(u_1)\,
\phi^{(\b)}_{m-\e_i,m-\e_i-\e_{j}}(u_2) +R^{\a\b}_{\b\a}(u_1-u_2)
\phi^{(\b)}_{m,m-\e_i}(u_1)\,
\phi^{(\a)}_{m-\e_i,m-\e_i-\e_{j}}(u_2)\no\\
&&\quad=W^{ij}_{ij}(u_1-u_2)
\phi^{(\a)}_{m-\e_{j},m-\e_{j}-\e_i}(u_1)\,
\phi^{(\b)}_{m,m-\e_{j}}(u_2)\no\\
&&\qquad\quad+W^{ji}_{ij}(u_1-u_2)
\phi^{(\a)}_{m-\e_{i},m-\e_{i}-\e_j}(u_1)\,
\phi^{(\b)}_{m,m-\e_{i}}(u_2),~~~i\neq j,~~\a\neq\b. \label{E-3}
 \eea
With the help of (\ref{f-fuction-1})-(\ref{f-fuction-3}) and  the
identity, \bea \sin(u)\,e^{\pm i\eta}+\sin(\eta)\,e^{\mp
iu}=\sin(u+\eta),\label{sin-id}\eea (\ref{E-1}) and (\ref{E-2})
can be proved case by case through tedious calculations.

(\ref{E-3}) is further divided into three
cases for $n\geq 3$:\bea &&|i-j|\geq 2,\,\,{\rm and}~ (i,j)\neq (1,n)~{\rm or\/}~(n,1),\no\\
&&|i-j|=1,\,\,{\rm and}~ (i,j)\neq (1,n)~{\rm or\/}~(n,1),\no\\
&&(i,j)=(1,n)~{\rm or\/}~(n,1),\no\eea and  one case for $n=2$:
\bea &&(i,j)=(1,n)~{\rm or\/}~(n,1).\no\eea Using
(\ref{f-fuction-1})-(\ref{f-fuction-3}) and after a
straightforward calculation, one can check (\ref{E-3}) for all of
the above cases. Therefore, we complete the proof of the
face-vertex correspondence relation (\ref{Face-vertex}) which
plays a key role to construct the K-matrices $K^{-}(u)$
(\ref{K-matrix}) and $K^{+}(u)$ (\ref{DK-matrix}).



\begin{thebibliography}{99}
\bibitem{Kor93} V.\,E. Korepin, N.\,M. Bogoliubov and A.\,G. Izergin,
{\it Quantum Inverse Scattering Method and correlation
Function\/}, Cambridge Univ. Press, Cambridge, 1993.
\bibitem{Skl88} E.\,K. Sklyanin, {\it Boundary conditions for
integrable quantum systems}, {\it J. Phys. \/} {\bf A 21} (1988),
2375.
\bibitem{Mez91} L. Mezincescue and R.\,I. Nepomechie, {\it Integrable open
spin chains with non-symmetric R-matrices}, {\it J. Phys.\/ } {\bf
A 24} (1991), L17.
\bibitem{Gho94} S. Ghoshal and A.\,B. Zamolodchikov, {\it Boundary s matrix and
  boundary state in two-dimensional integrable quantum field
  theory}, {\it Int. J. Mod. Phys.\/} {\bf A 9} (1994), 3841
  [{\tt hep-th/9306002}].
\bibitem{Dev93} H.\,J. de Vega and A.\,G. Ruiz, {\it Boundary K-matrices for the
six-vertex and $n(2n-1)$ $A_{n-1}$ vertex models}, {\it J.
Phys.\/} {\bf A 26} (1993), L519.
\bibitem{Aba95} J. Abad and M. Rios, {\it Non-diagonal solutions
to reflection equations in $su(n)$ spin chains}, {\it Phys.
Lett.\/} {\bf B 352} (1995), 92.
\bibitem{Gan99} G.\,M. Gandenberger, {\it New non-diagonal
solutions to the $a^{(1)}_n$ boundary Yang-Baxter equation}, {\tt
e-print: hep-th/9911178}.
\bibitem{Lim02} A. Lima-Santos,
{\it $A^{(1)}_{n-1}$ reflection K-matrices},  {\bf B 644\/}
(2002), 568.
\bibitem{Fan98} H. Fan, B.\,Y. Hou, G.\,L. Li and K.\,J. Shi, {\it
A new solution to the reflection equation for the $Z_n$ symmetric
Belavin model},  {\it Phys. Lett.\/} {\bf A 250\/} (1998), 79.
\bibitem{Yan041}  W.\,-L. Yang and R. Sasaki, {\it Solution of the dual reflection
  equation for $A^{(1)}_{n-1}$ SOS model},
{\it J. Math. Phys.} {\bf 45} (2004), 4301 [{\tt hep-th/0308118}].
\bibitem{Mez98}  L. Mezincescue and R.\,I. Nepomechie, {\it Fractional-spin
integrals of motion for the boundary sine-Gordon model at the free
fermion point}, {\it Int. J. Mod. Phys.\/ } {\bf A 13} (1998),
2747.
\bibitem{Del01} G.\,W. Delius and N.\,J. Mackay, {\it Quantum
group symmetry in sine-Gordon and affine Toda field theories on
the half-line}, {\it Commun. Math. Phys.\/} {\bf 233\/} (2003),
173 [{\tt hep-th/0112023}].
\bibitem{Arn03} D. Arnaudon, J. Avan, N. Crampe, A. Doikou, L.
Frappat and E. Ragoucy, {\it Classification of reflection matrices
related to (super-) Yangians and application to open spin chain
models}, {\it Nucl. Phys.\/} {\bf B 668} (2003), 469; {\it General
boundary conditions for the $sl(N)$ and $sl(M|N)$ open spin
chains}, {\tt e-print: math-ph/0406021}.
\bibitem{Doi04} A. Doikou, {\it From affine Heck  algebras to
boundary symmetries},  {\tt e-print: math-ph/0409060}.
\bibitem{Bel81} A. Belavin, {\it Dynamical symmetry of integrable quantum
systems}, {\it Nucl. Phys.\/} {\bf B 180} (1981), 189.
\bibitem{Yan043} W.\,-L. Yang, Y.\,-Z. Zhang and M. Gould,
{\it Exact solution of the XXZ
  Gaudin model with generic open boundaries}, {\it Nucl. Phys.\/}
  {\bf B 698} (2004), 503 [{\tt hep-th/0411048}].
\bibitem{Yan042} W.\,-L. Yang and Y.\,-Z. Zhang, Exact
solution of the $A^{(1)}_{n-1}$ trigonometric vertex model with
non-diagonal open boundaries, {\tt e-print: hep-th/0411190}.
\bibitem{Yan04} W.\,-L. Yang and R. Sasaki, {\it Exact solution of $Z_n$ belavin model
  with open boundary condition}, {\it Nucl. Phys.\/}
{\bf B 679} (2004), 495 [{\tt hep-th/0308127}].
\bibitem{Min01} M. Mintchev, E. Ragoucy and P. Sorba, {\it Spontaneous symmetry breaking in the
$gl(N)$-NLS hierarchy on the half line}, {\it J. Phys.\/} {\bf A
34} (2001), 8345.
\bibitem{Gal04} W. Galleas and M.\,J. Martins, {\it Solution of the $SU(N)$ vertex model
with non-diagonal open boundaries}, {\tt e-print:
nlin.SI/0407027}.
\bibitem{Che80} I.\,V. Cherednik, {\it On a method of constructing factorized
$S$-matrices in terms of elementary functions},  {\it Theor. Mat.
Fiz\/} {\bf 43} (1980), 117.
\bibitem{Per81} J.\,H.\,H. Perk and C.\,L. Schultz, {\it New families of commuting
transfer matrices in $q$-state vertex models}, {\it Phys. Lett.\/}
{\bf A 84} (1981), 407.
\bibitem{Baz91} V.\,V. Bazhanov, R.\,M. Kashaev, V.\,V. Mangazeev
and Yu.\,G. Stroganov, {\it $(Z_N\times)^{n-1}$ generalization of
the chiral potts model}, {\it Commun. Math. Phys.\/} {\bf 138}
(1991), 393.

\bibitem{Jim87} M. Jimbo, T. Miwa and M. Okado, {\it Solvable lattice models
whose states are dominant integral weights of $A^{(1)}_{n-1}$},
{\it Lett. Math. Phys.\/} {\bf 14} (1987), 123; {\it Local state
probabilities of solvable lattice modles: $A^{(1)}_{n-1}$ family},
{\it Nucl. Phys.\/} {\bf B 300} (1988), 74.
\bibitem{Hou03} B.\,Y. Hou, R. Sasaki and  W.\,-L. Yang, {\it Algebraic
bethe ansatz for  the elliptic quantum group $e_{\tau,\eta}(sl_n)$
and its  applications},  {\it Nucl. Phys.\/} {\bf B 663} (2003),
467 [{\tt hep-th/0303077}].


\end{thebibliography}
\end{document}